\documentclass[conference]{IEEEtran}
\IEEEoverridecommandlockouts
\usepackage{cite}
\usepackage{caption}
\usepackage{amsmath,amssymb,amsfonts}
\usepackage{algorithmic}
\usepackage{graphicx}
\usepackage{textcomp}
\usepackage{xcolor}
\usepackage{bm}
\usepackage{subfigure}
\def\BibTeX{{\rm B\kern-.05em{\sc i\kern-.025em b}\kern-.08em
    T\kern-.1667em\lower.7ex\hbox{E}\kern-.125emX}}
\begin{document}

\title{Towards a Better Microcredit Decision\\
}
\author{
\IEEEauthorblockN{ Mengnan Song}
\IEEEauthorblockA{\textit{360 DigitalTech} \\
Beijing, China \\
songmengnan-jk@360shuke.com}
\and
\IEEEauthorblockN{ Jiasong Wang}
\IEEEauthorblockA{\textit{360 DigitalTech} \\
Beijing, China \\
wangjiasong-jk@360shuke.com}
\and
\IEEEauthorblockN{ Suisui Su}
\IEEEauthorblockA{\textit{360 DigitalTech} \\
Beijing, China \\
susuisui@360shuke.com}
}

\maketitle

\begin{abstract}
Microcredit refers to personal oriented loans for consumption, which is mainly characterized by small loans, medium-short term and lacking collateral, and serves high-risk populations who are highly likely to be rejected by traditional financial institutions. A crucial challenge that the leading platforms are facing is to assess the creditworthiness of applicants and decide whether or not to grant credit. Therefore, credit scorecard emerges and has drawn broad attention, as it takes an important role in the field of risk management. However, the parameters of scorecard are usually estimated using a sample set that excludes rejected applicants, which may prove biased when applied to all applicants.

Reject inference comprises techniques to infer the possible repayment behavior of rejected cases. In this paper, we model credit in a brand new view by capturing the sequential pattern of interactions among multiple stages of loan business to make better use of the underlying causal relationship. Specifically, we first define 3 stages with sequential dependence throughout the loan process including credit granting(AR), withdrawal application(WS) and repayment commitment(GB) and integrate them into a multi-task architecture. Inside stages, an intra-stage multi-task classification is built to meet different business goals. Then we design an Information Corridor to express sequential dependence, leveraging the interaction information between customer and platform from former stages via a hierarchical attention module controlling the content and size of the information channel. In addition, semi-supervised loss is introduced to deal with the unobserved instances. The proposed multi-stage interaction sequence(MSIS) method is simple yet effective and experimental results on a real data set from a top loan platform in China show the ability to remedy the population bias and improve model generalization ability.
\end{abstract}


\section{Introduction}
Consumer financial behavior involving obtaining cash flow from platforms within a specified period to meet consumption needs by paying interest expense. As a new financial tool, Micro-finance grants a small credit line to low-income populations who often lack historical credit behavior because of rejection by traditional financial institutions. In recent years, online lending business in China has achieved tremendous development and a group of leading platforms serving a huge customer base has emerged. One of the challenges in this so called Inclusive Finance is to assess the creditworthiness of applicants and decide whether or not to grant credit, and credit scoring method predicting whether a borrower would default or not takes an important role in the field of risk management\cite{thomas2017credit}. Since its introduction as a statistic method in the 1950s, credit scoring has drawn broad attention and transitioned to machine learning method as a standard form of binary classification using logistic regression or boosting tree which is more popular currently with adequate features as input\cite{mester1997s}\cite{chye2004credit}. As even a minor improvement can prevent dramatic loss, the further research of credit scoring is necessary and lies in 2 directions, namely feature engineering and methodology. The main job of engineers in the industry is to explore predictive features to enhance the model performance, while \cite{song2020effective} proposes an automated feature engineering framework highlighting the role of AutoML in the modeling work. In terms of methodology, coupled with development in computing power and growing complexity of business problems, many other advanced alternatives have gained attention including reject inference.

In general, credit scoring model is developed based on the accepted applications of which the repayment behavior in the past is known to identify customers as good or bad, called GB model, assuming that the samples used for model training can represent the overall population\cite{greene1998sample}. However in practice, this assumption does not hold, because the amount of rejected credit requests is much larger than that of accepted ones, especially in Microcredit business and we are unable to observe the repayment behavior of the declined request. It should be pointed out that the AR model based on entire sample space predicting the acceptance probability of credit requests also contains certain information that should not be waste. Since the credit model is typically designed to assess the whole through-the-door population, the sample selection bias turns to be a serious issue and reject inference techniques are one way to address this concern\cite{nguyen2016reject}. Reject inference refers to inferring the possible loan repayment outcomes of rejected samples with the associated credit model established based on both the accepted and rejected observations\cite{anderson2019using}. Thus the credit model is representative of the entire loan applications. There is a vast research on reject inference about traditional statistical methods, yet little published work using machine learning techniques. Ref.\cite{mancisidor2020deep} presents a research overview on reject inference in risk management from which we can see that the semi-supervised learning is popular in recent years. However, all these methods are explored from the perspective of academic by making assumptions to simplify the problem, therefore, unable to meet the industry needs. Besides, there exists a sequential dependency between AR model and GB model, but the role of AR model is ignored, not to mention the WS model predicting the probability of withdrawal or silence for users after granted a credit line.

\begin{figure}
\centering
\subfigure[Deep Exploration]{\label{}\includegraphics[scale=0.45]{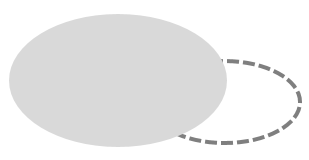}}
\subfigure[Shallow Exploration]{\label{}\includegraphics[scale=0.45]{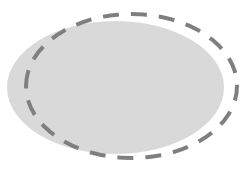}}
\centering
\subfigure[Dual Task Model Structure]{\label{}\includegraphics[scale=0.25]{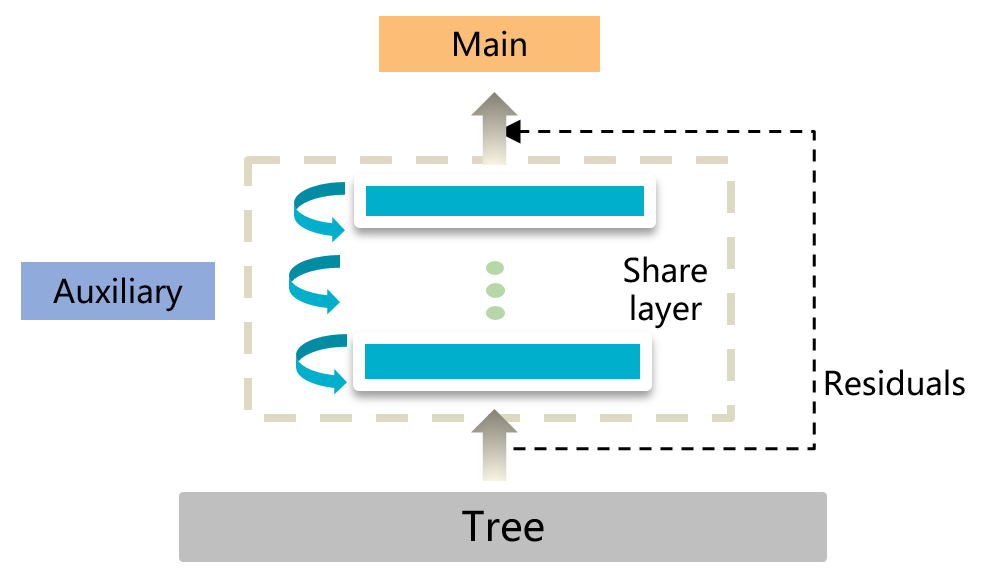}}
\caption{Model Stability Exploration}
\label{dual-task}
\end{figure}

Meanwhile, in the field of recommendation, multi-task learning brings a new perspective for similar sample bias problem during estimating post-click conversion rate with the idea of incorporating auxiliary task to improve the learning of primary target\cite{wang2019deep}\cite{ma2018entire}\cite{xi2021modeling}. Our previous preliminary explorations on model stability demonstrate the effectiveness of multi-task learning. A real business scenario is identifying relatively good customers with lower risk from rejected applicants to improve the acceptance rate and serve more users. Usually we will sample a subset of poor users from the whole accepted user pool based on some specific strategies and then the sampled users will be utilized to train a GB model. However, the online performance of this model is not stable and will decay quickly. As shown in Fig. \ref{dual-task}(a), the poor users filtered out according to certain rules is only a small local part of the entire space, leading to the model taking a relatively deep exploration along a certain direction based on the original decision boundary. While a more reasonable way might be exploring uniformly around the decision boundary with a small step size to maintain the stability of the model, as shown in Fig. \ref{dual-task}(b). To achieve this goal, we design a mixture model combining tree based model and neural network with the AR target as an auxiliary task to balance the GB decision, which is called dual task model and  has been running stably online for a year. Another work tests more targets and views them all equally, as there exists different business goals in WS and GB stages. Different targets are complementary and can promote each other bringing a 2\% increase in AUC which is a common model evaluation indicator. A important observation is that instead of being independent or mutually inhibitive, multiple tasks are generally correlated and possess significant potential of being in synergy\cite{wang2019deep} and our practice shows the value of standard multi-task learning, but this formulation also does not take the causal relationship among targets into consideration.

In this paper, we propose a MSIS method to make better use of the AR and WS models by capturing the sequential pattern of interactions among multiple stages of loan business based on multi-task architecture. The AR and WS models are auxiliary tasks, while the GB model predicting default probability is our major target. To the best of our knowledge, this is the first work that leveraging platform decision and user willingness to help infer the status of declined applicants. Our contributions throughout this paper are as follows.

First, we define 3 stages with sequential dependence within the loan process that are AR, WS and GB and then integrate them into a multi-task architecture. Sample selection mechanism exists in both the AR and WS stages, as applicants may be refused to grant credit or keep silent after granted a credit line. To meet different business goals, a intra-stage multi-task classification is built, for the definition of label is ambiguous and different business goals suggest different ways in real business. They can of course be viewed as multiple independent binary classifications, but our practice reveals that training them together through a multi-task style can promote each other.

Second, we design an Information Corridor to explicitly express the sequential dependency among multiple stages. A direct connection is built on the upper part of the multi-task architecture which is called tower to transfer information from the former stage to the latter. A hierarchical attention mechanism consisting of intra-stage attention and inter-stage attention is designed to learn to control the content and size of the information channel during training. The intra-stage attention filters and aggregates information from different targets within stage, while the inter-stage attention transfers and redistributes information from the former stage to different targets in the latter stage.

Third, semi-supervised loss is applied to deal with the unlabeled data. As mentioned above, the selection mechanism exists in both the AR and WS stages and the repayment behavior of users can not be observed in 2 situations that are rejected to grant credit by platform and keeping silent after granted credit. Semi-supervised method can naturally handle missing data and a common practice is to add a regularization term with an entropy constraint to the objective function.

The remainder of this paper is organized as follows. Literatures of related work are described in section 2. Section 3 introduces the MSIS method and its modules. In section 4, groups of experiments are implemented to evaluate the efficiency of the proposed method. Conclusions and future works are summarized in section 5.

\section{Related Works}
Reject inference is the process of estimating default risk for loan applications that are rejected under the current decision system to mitigate the bias problem. The reject inference is formulated as a problem of learning with missing data which can be classified into missing at random(MAR) and missing not at random(MNAR) in the sense of \cite{little2019statistical}\cite{rubin1976inference}. MAR means that the probability distribution of default is the same whether applications are accepted or rejected, indicating that the decision system of lending platform is ineffective. Ref.\cite{feelders2000credit} includes the rejects in the estimation process by using a mixture model formulation with EM method estimating the model parameters. However, as the risk management of leading platforms is performing well, the MAR assumption does not correspond to reality. Under the MNAR assumption, according to \cite{siddiqi2012credit}, reject inference is invalid when the confidence level is very high that all rejected ones are bad or very low that credit is granted randomly.

The early research consider mainly on statistical methods, including reweighing, extrapolation and Heckman method. Reweighting involves weighting accepted samples to scale up to the total population using the ratio between the number of approved samples and declines in a group\cite{hsia1978credit}\cite{crook2004does}\cite{banasik2007reject}. As AR model scores represent the probability of being accepted, the grouped weightings are then replaced by the AR model scores. The major advantage of reweighting is simplicity. However, this method only brings improvement by chance for lacking knowledge of the tendency to become delinquent of the rejects. Ref.\cite{crook2004does}\cite{parnitzke2005credit} demonstrate that reweighting does not really promote the performance of the GB model. Extrapolation refers to assign a pseudo label to the rejected samples using the same GB model and then re-train the GB model with all accepted and rejected samples\cite{ash2002best}. The default probability of rejected samples based on the original GB model will be used to simulate outcomes either by Monte Carlo parceling or fuzzy augmentation\cite{montrichard2007reject}. Ref.\cite{parnitzke2005credit} shows that extrapolation leads to significant improvement in the model and \cite{zeng2014rule} claims that fuzzy augmentation performs better than others. Heckman's two-step model simultaneously models the AR and GB stages assuming that the error terms are bivariate normally distributed. \cite{banasik2003sample} finds that the Heckman method can get an improvement using a proprietary data set, but the gain is relatively small. However other researchers demonstrate that Heckman model does not work well\cite{puhani2000heckman}\cite{chen2012bound}.

Machine learning techniques like unsupervised and semi-supervised methods are popular in recent years, as they can naturally handle unlabeled data together with labeled data. Ref.\cite{shen2020three} proposes a three-stage reject inference learning framework based on unsupervised transfer learning and three-way decision theory. It learns a higher level representation of data based on both accepted and rejected samples with a three-way decision theory to select a subset of rejected samples to avoid negative transfer problem. The rejected samples are also filtered in \cite{kozodoi2019shallow} with isolation forest and then self-learning is employed to estimate model parameters using all accepted samples and the selected rejected ones. Due to the good performance and simplicity, SVM is also a commonly used semi-supervised source tool. Ref.\cite{li2017reject} directly applies S3VM to simultaneously model the accepted and rejected applicants and \cite{tian2018new} further develops a kernel-free fuzzy quadratic surface SVM addressing the issues of hyper-parameter tuning and scalability. Also by means of self-learning, \cite{maldonado2010semi} adopts SVM as base classifier with a modification in order to incorporate the assumption that the rejected loans have a higher risk. Inspired by ensemble ideas, \cite{kim2019ensemble} combines label propagation and TSVM, as label propagation learns similar features in the same class and TSVM learns different features among classes. Besides, \cite{mancisidor2020deep} combining gaussian mixtures and auxiliary variables in a semi-supervised framework with generative models infers the unknown credit worthiness of the rejected applications by exact enumeration of the two possible outcomes of the loan.

Meanwhile, in the field of recommendation, multi-task learning brings a new perspective for similar sample bias problem during estimating post-click conversion rate by better expressing the underlying correlation among targets which can be implemented in deep networks by hard parameter sharing\cite{1993Multitask}, soft parameter sharing\cite{yang2016trace} and cross-stich networks\cite{2016Cross}. The dependence relationship among tasks is also taken into consideration, as clear-cut causal relationship is hard-coded in to the model in \cite{2018Entire} and less obvious relationship that can not be described directly is captured by Bayesian network\cite{2019Deep} and attention mechanism\cite{xi2021modeling}.

\section{The Proposed MSIS Method}
The proposed method is based on a hierarchical structure with inter-stage level and intra-stage level manifested in two aspects including tasks and information flow. For tasks, the inter-stage tasks represent different stages and the intra-stage tasks refer to different business goals inside stages. In terms of information flow, the inter-stage flow involve information transferred from the former stage to the latter, while the intra-stage flow represent information aggregation inside stages. In this section, we discuss the detail of the proposed method.

\subsection{Multi-task Learning with 3 Business Stages}
\begin{figure}[htbp]
\centerline{\includegraphics[scale=0.3]{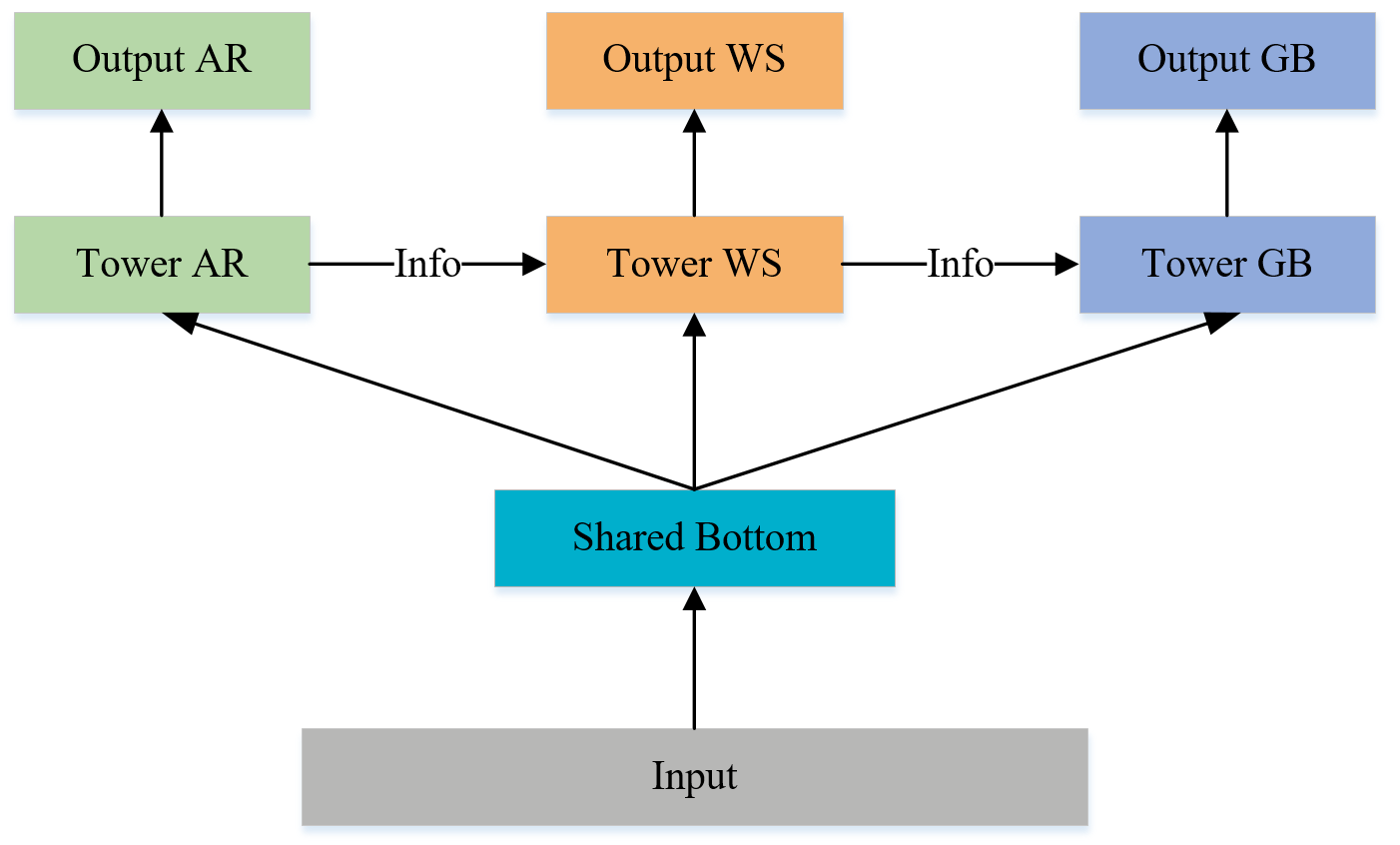}}
\caption{MSIS Model Structure}
\label{multi-task}
\end{figure}

Given the input feature \bm{$x$} which is collected when user submits a credit application ensuring that there is no information leakage in terms of time line, generally, there exists 3 stages throughout the loan process that are AR, WS and GB. Firstly, after registration, the customer need submit a credit application to the platform to obtain a certain credit line and the lending platform should decide whether or not to grant credit according to its risk management system. After accepted by the lending platform and obtaining a credit line, users may or may not make withdrawal requests under their willingness and financial status. Finally, assuming all withdrawal requests will be approved by platform, loan users should pay back the loan according to the agreement. Obviously, sample selection mechanism exists in both the AR and WS stages. These 3 stages can be naturally integrated into a multi-task architecture with causal relationship between adjacent stages. The overall model structure is shown in Fig. \ref{multi-task} with input layer, shared layer, tower layer and Information Corridor. Shared layer and tower layer capture the common features and specific features from different tasks respectively. Information Corridor models the dependence relationship among stages.

Let $\theta $ denotes the model parameters and $t$ denotes the stage. The 3 stages are $t_{AR}$, $t_{WS}$ and $t_{GB}$. Based on the above architecture, from a probabilistic point of view, according to the Bayes' rule, we have

\begin{equation}
\begin{split}
&P(t_{AR},t_{WS},t_{GB}|\bm{x},\theta) \\
&=P(t_{AR})*P(t_{WS}|t_{AR})*P(t_{GB}|t_{AR},t_{WS}) \\
&=P(t_{AR})*P(t_{WS}|t_{AR})*P(t_{GB}|t_{WS})
\end{split}
\end{equation}
The overall loss function can be written as:
\begin{equation}
\begin{split}
&L(\bm{x},\theta) \\
&=w_{AR}  L(t_{AR})+w_{WS}  L(t_{WS},t_{AR})+w_{GB}  L(t_{GB},t_{WS}) \\
\end{split}
\end{equation}
with different weights for each term to control the relative importance of stages.

In the AR stage, the only outcome is either acceptance or rejection, which is clearly a simple binary classification problem. However, it is not the same case in the other two stages where the definition of labels can be ambiguous and the intra-stage multi-task classification is built for different definitions pursuing different business goals. In the WS stage, the time gap between credit granting and withdrawal can express the willingness and financial status of user to a certain extent. So labels in this stage can be designed as users whether or not will make a withdrawal requests within 30 days and 90 days. In the GB stage, as the installment business usually involves multiple repayments, labels are designed as users whether or not will default within 1 term, 3 terms and 6 terms. They can of course be viewed as multiple independent targets, but our previous works reveal that training them together through a multi-task style can bring complementary effects. The intra-stage multi-task structure in the GB stage is shown in the right part of the Fig. 3, which is similar with the structure of the WS stage.

\begin{figure*}[htbp]
\centering{\includegraphics[scale=0.4]{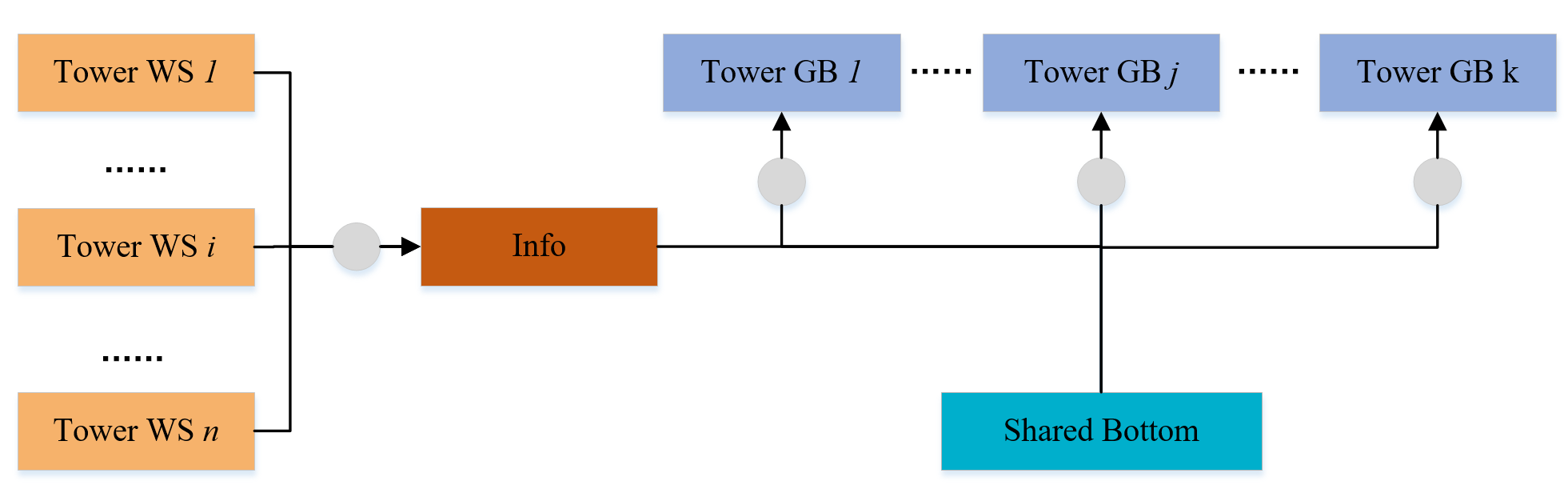}}
\caption{Information Corridor.}
\label{information-Corridor}
\end{figure*}

\subsection{The Information Corridor}

Information Corridor is designed to explicitly express sequential dependency among multiple stages by introducing a direct connection on the tower part of the multi-task architecture to transfer information from the former stage to the latter. Let $\bm{e}_{ou}$ denotes the output of the information channel in the former stage and $\bm{e}_{in}$ denotes the input of the information channel in the latter stage, the information transforming is defined as:
\begin{equation}
\bm{e}_{in}=f(\bm{e}_{ou})
\end{equation}
$k_{ou}$ and $k_{in}$ is the dimension of $\bm{e}_{ou}$ and $\bm{e}_{in}$ respectively with $\bm{e}_{ou}\in \mathbb{R}^{k_{ou}}$ and $\bm{e}_{in}\in \mathbb{R}^{k_{in}}$. $f()$ is a non-linear function to learn to retain useful information and discard redundant information between stages. Then we describe the information flow between stages as two steps including aggregation and redistribution. In the former stage, information from different intra-stage targets is filtered and aggregated into $\bm{e}_{ou}$. Then after the information transformation, $\bm{e}_{in}$ is redistributed to intra-stage targets of the latter stage and each target shall filter the information of $\bm{e}_{in}$ according to its own representation. Inspired by \cite{xi2021modeling}\cite{2017Attentional}, a hierarchical attention mechanism consisting of intra-stage attention and inter-stage attention is designed to learn to control the content and size of the information channel during training process.

Taking the WS and GB stages as example with information transferring from WS to GB stage, let $\bm{h}_{WS}^{1}$ and $\bm{h}_{WS}^{2}$ denotes the tower representations of targets in WS stage. The output $\bm{e}_{ou}$ is calculated as:

\begin{equation}
\bm{e}_{ou}=\sum_{m}\alpha_{m}g_{3}(\bm{h}_{WS}^{m})
\end{equation}
$k_{WS}^{1}$ and $k_{WS}^{2}$ is the dimension of $\bm{h}_{WS}^{1}$ and $\bm{h}_{WS}^{2}$ respectively with $\bm{h}_{WS}^{1}\in \mathbb{R}^{k_{WS}^{1}}$ and $\bm{h}_{WS}^{2}\in \mathbb{R}^{k_{WS}^{2}}$. $g_{3}()$ is a non-linear function to project the input vector to a new vector representation. $\bm{\alpha}$ is the weight vector of targets in WS stage and is represented as:
\begin{equation}
\alpha_{m}=\frac{exp(\hat{\alpha}_{m})}{\sum_{m}exp(\hat{\alpha}_{m})}
\end{equation}
\begin{equation}
\hat{\alpha}_{m}=\frac{<g_1(\bm{h}_{WS}^{m}),g_2(\bm{h}_{WS}^{m})>}{\sqrt{k_{WS}^{m}}}
\end{equation}
where $<.,.>$ represents the dot product. The intra-stage attention filters and aggregates information from the towers of the two targets in the WS stage with a weight vector.

In the GB stage, let $\bm{h}_{GB}^{1}$ and $\bm{h}_{GB}^{2}$ denotes the tower representations, $\hat{\bm{h}}_{GB}^{1}$ and $\hat{\bm{h}}_{GB}^{2}$ denotes the new tower representations that are the fusions of $\bm{e}_{in}$ and $\{\bm{h}_{GB}\}$. $\{\hat{\bm{h}}_{GB}\}$ are the top feature layers and are used as the direct input of classification layer. The fusion is described as:
\begin{equation}
\hat{\bm{h}}_{GB}^{1}=\beta_{1}g_{3}^{'}({\bm{e}_{in}})+\beta_{2}g_{3}^{'}({\bm{h}_{GB}^{1}})
\end{equation}
$g_{3}^{'}()$ is also a non-linear function. $\bm{\beta}$ is the weight vector like the above $\bm{\alpha}$ and the calculation of $\bm{\beta}$ is also similar with $\bm{\alpha}$. The inter-stage attention redistributes the information $\bm{e}_{in}$ from the WS stage by fusing $\bm{e}_{in}$ and each tower representation with a weight vector. The calculation of of $\hat{\bm{h}}_{GB}^{1}$ can be applied to $\hat{\bm{h}}_{GB}^{2}$ and other intra-stage targets in the GB stage.

There is no intra-stage attention, as only one target in the AR stage. In the Information Corridor we extend the self-attention mechanism to a hierarchical structure to realise fine-grained control over information channels. A series of non-linear functions $\{g\}$ and $\{g^{'}\}$ represent the feed-forward networks and parameter sharing does not bring improvement in out experiments.

\subsection{Semi-supervised Loss}
As we have discussed above, selection mechanism exists in both the AR and WS stages, producing unlabeled samples whose repayment behaviors are unobserved including the rejected applicants in the AR stage and silent users in the WS stage.

A simple and efficient semi-supervised method that can naturally handle the unlabeled data is introduced by adding a regularization term with an entropy constraint to the objective function. For example, let $t_{GB}^{1}$ denotes one of the targets which is a typical binary classification with $p_{GB}^{1}$ as the output probability and $y_{GB}^{1}$ as the true label in the GB stage. The corresponding entropy is defined as:
\begin{equation}
H(p_{GB}^{1})=-p_{GB}^{1}*logp_{GB}^{1}-(1-p_{GB}^{1})*log(1-p_{GB}^{1})
\end{equation}
For unlabeled data, entropy constraint regularization is expressed as:
\begin{equation}
\Omega(t_{GB}^{1})=-\sum_{i=1}^{m} \{p_{GB}^{1,m}*logp_{GB}^{1,m}+(1-p_{GB}^{1,m})*log(1-p_{GB}^{1,m})\}
\end{equation}
where m is the number of unlabeled instances. Because the total number of labeled data and unlabeled data is quite different and the training balance between them is quite important for the network performance, the overall loss function is
:
\begin{equation}
L(t_{GB}^{1})=L_{sup}(t_{GB}^{1})+\gamma_{GB}^{1} * \Omega(t_{GB}^{1})
\end{equation}
The above loss function is also applied to other targets.

\section{EXPERIMENTS}
The proposed method is evaluated with a real data from a top loan platform in China and compared with stage-of-the-art methods.

\subsection{Dataset}
The dataset contains about 4 million credit applications, including about 400,000 loans in the GB stage. All experiments are repeated 5 times and averaged results are reported. According to the common practice of financial industry, test data is defined as Out of Time samples, in which the training and validation sets do not coincide in time. The online decision system of this platform is constantly adjusted to adapt to the internal cognition and external environment. There is a large policy adjustment between the collection of training and test data, so the fitness to test data can be used to evaluate the generalization ability. In the dataset, each sample has 32-dimensional features including user profile and credit history. For the three stages of AR, WS and GB, six labels are defined with 2 in WS stage and 3 in GB stage to meet different business goals as follows:
\begin{itemize}
  \item \textbf{label\_credit}: Whether the user is rejected or approved by the risk management system, in AR stage.
  \item \textbf{label\_draw\_30}: Whether the user makes a withdrawal request within 30 days after granted credit, in WS stage.
  \item \textbf{label\_draw\_90}: Whether the user makes a withdrawal request within 90 days after granted credit, in WS stage.
  \item \textbf{label\_mob1}: Whether the user is overdue for more than 30 days in the \textbf{first} repayment period, in GB stage.
  \item \textbf{label\_mob3}: Whether the user is overdue for more than 30 days in the \textbf{third} repayment period, in GB stage.
  \item \textbf{label\_mob6}: Whether the user is overdue for more than 30 days in the \textbf{sixth} repayment period, in GB stage.
\end{itemize}

\subsection{Competitive Methods}
Our work is compared with the following methods:
\begin{itemize}
  \item \textbf{Xgboost}\cite{chen2016xgboost}: Xgboost is a tree based model with gradient boosting learning algorithm. It is widely used in all walks of life and has achieved great success especially in the financial field. Almost all major models of leading platform are based on Xgboost. We take it as the benchmark model.
  \item \textbf{MLP}: Multi-Layer Perceptron is a fully connected neural network model with multiple layers.
  \item \textbf{MLP-Entropy-Reg}: Entropy regularization is a simple and efficient semi-supervised method that can naturally handle the unlabeled data by adding a regularization term with an entropy constraint to the objective function. MLP with entropy regularization can broadly represent semi-supervised methods.
  \item \textbf{Self-Learning}: Self-Learning is also a common semi-supervised learning method, which assigns labels to rejected samples in accordance with the model trained by only labeled data and then retrains the model with all samples. This procedure is repeated until certain criteria is met. In our experiments, this method is implemented based on the Xgboost.
  \item \textbf{MMOE}\cite{ma2018modeling}: MMOE is a multi-task learning model splitting the shared  layers into subnetworks and using  gating networks to organize subnetworks for different tasks.
  \item \textbf{AITM}\cite{xi2021modeling}: The Adaptive Information Transfer Multi-task framework models the sequential dependence among audience multi-step conversions. But it supports only 1 target in each conversation step and can not deal with unlabelled data.
\end{itemize}

\subsection{Performance Comparison}
\begin{table*}[htbp]
\caption{ The AUC performance (mean$\pm$std) on our business dataset. The Gain means the mean AUC improvement compared with the Xgboost. We use '*' to indicate the improvement of MSIS over the best performance from the baselines is significant based on paired t-test at the significance level of 0.01.}
\begin{center}
\begin{tabular}{|c|c|c|c|c|c|c|}
\hline
\textbf{}&\multicolumn{3}{|c|}{\textbf{AUC}} &\multicolumn{3}{|c|}{\textbf{Gain}}\\
    \cline{2-7}
    \textbf{Model} & \textbf{\textit{label\_mob1}}& \textbf{\textit{label\_mob3}}& \textbf{\textit{label\_mob6}}& \textbf{\textit{label\_mob1}}& \textbf{\textit{label\_mob3}}& \textbf{\textit{label\_mob6}} \\
    \hline
        Xgboost& 0.5880$\pm$0.0068& 0.5976$\pm$0.0083& 0.6030$\pm$0.0159&  -&  -&   - \\
        MLP& 0.5585$\pm$0.0043& 0.5661$\pm$0.0051& 0.5791$\pm$0.0031& -0.0295& -0.0315& -0.0239  \\
        MLP-Entropy-Reg& 0.5655$\pm$0.0069&	0.5744$\pm$0.0038&	0.5867$\pm$0.0056&	-0.0224& 	-0.0232& 	-0.0163 \\
        Self-Learning&	0.5973$\pm$0.0137&	0.6035$\pm$0.0046& 0.5957$\pm$0.0046&	+0.0094& +0.0059& -0.0073 \\
        MMOE& 0.5838$\pm$0.0133& 0.6011$\pm$0.0016& \underline{0.6082$\pm$0.0016}& -0.0042& +0.0036& +0.0052  \\
        AITM& \underline{0.6049$\pm$0.0030}& \underline{0.6064$\pm$0.0018}& 0.6021$\pm$0.0013& +0.0169& +0.0089& -0.0009  \\
        MSIS& \textbf{0.6103$\pm$0.0223$^*$}& \textbf{0.6228$\pm$0.0252$^*$}& \textbf{0.6229$\pm$0.0199$^*$}& +\textbf{0.0223}& +\textbf{0.0252}& +\textbf{0.0199} \\
    \hline
\end{tabular}
\label{compare}
\end{center}
\end{table*}
The main idea of our work is to incorporate auxiliary tasks to improve the learning of primary target by introducing the sequential dependence among tasks. The auxiliary tasks are AR and WS models, while the primary target is GB model. In this subsection, we compare the performance of the proposed model with existing state-of-the-art methods on three labels in GB stage. We record the mean and standard deviation of AUC of the above mentioned methods by repeating the experiment five times. Taking xgboost model as baseline, the gain of AUC is calculated, and the results are shown in Table.\ref{compare}. Through the analysis of the results, the major findings can be summarized as follows:

\begin{itemize}
  \item MLP works poorly on these three labels, as the results are 2.95\%, 3.15\% and 2.39\% lower than that of the xgboost model respectively which is consistent with our practical experience. In the field of risk management, features are usually developed by experts presented as tabular data. This structured data is unable to exert the advantages of deep learning in feature extraction. So as the same binary classification model, the MLP performs worse than xgboost.
   \item Semi-supervised learning has a certain effect. Entropy regularization and self-train are two commonly used semi-supervised methods and it is observed that they can alleviate sample selection bias to some extent and improve the model performance. Although MLP-Entropy-Reg still does not achieve positive gain compared with the baseline, it outperforms the original MLP. Compared with original xgboost, self-train brings some positive gains in terms of label\_mob1 and label\_mob3.
  \item Compared with single-task model, the multi-task methods including MMOE, AITM and MSIS can always bring benefits which proves different business targets promote each other and the information from the  stages benefit the decision of GB model.
  \item The performance of MMOE is not as good as that of AITM and MSIS which reveals that the causal relationship among stages should not be ignored.
  \item The proposed method outperforms AITM for all the three labels by 0.54\%, 1.64\% and 2.08\% respectively, especially for the long term target, which indicates that the hierarchical attention mechanism realises fine-grained control over information channels and semi-supervised learning method make efficient use of unlabeled data.
  \item Compared with the benchmark model, the proposed MSIS method achieves the most gain and surpasses other state-of-the-art methods which demonstrates its generalization ability.
\end{itemize}

\subsection{Ablation Study}
\begin{figure*}[htbp]
\centering
\subfigure[ Effectiveness of each module ]{\label{}\includegraphics[scale=0.45]{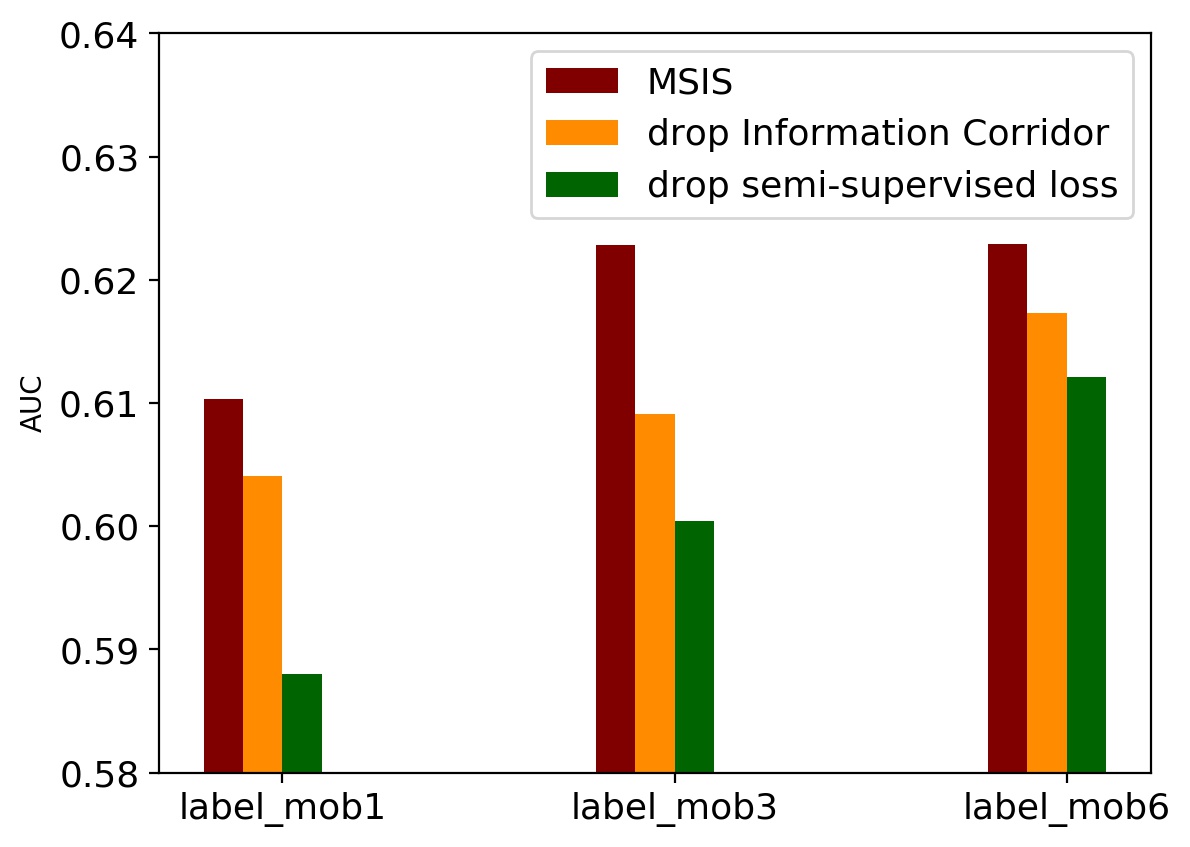}}
\subfigure[ Exploration of stage numbers ]{\label{}\includegraphics[scale=0.45]{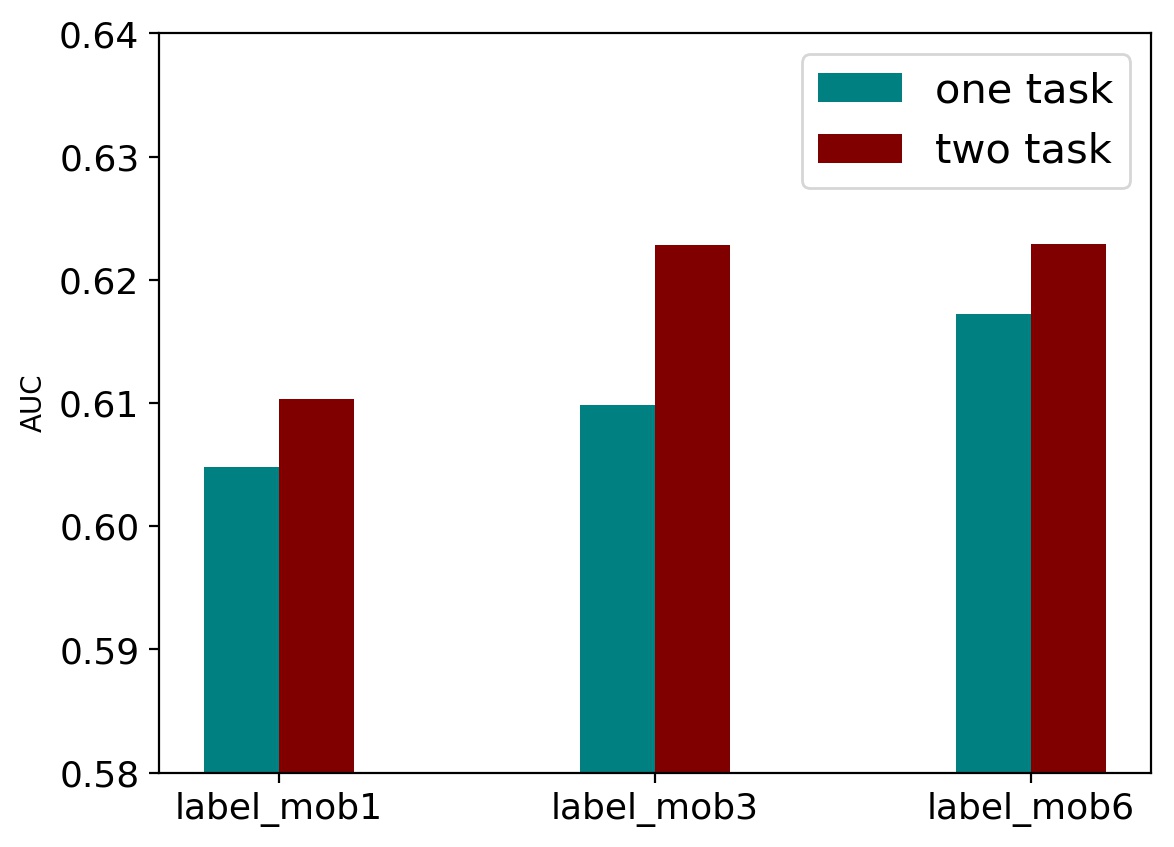}}
\caption{The mean AUC performance in Ablation Study}
\label{ablation}
\end{figure*}

We perform a ablation study on the modules of intra-stage targets, semi-supervised loss and business stages. One of the modules is removed in each time, results are shown in Fig. \ref{ablation}.

First, we study the effectiveness of the intra-stage targets. Assuming only one single target remaining within each stage, the AUC will decrease by 0.62\%, 1.37\%, 0.56\% for the 3 default labels respectively. This shows that the multiple targets inside stage promote each other and the hierarchical attention mechanism realises fine-grained control over information flow.

Then, we study the benefits of semi-supervised loss. We directly remove this part from the loss function, that is, only use labeled samples for training, and the AUC of the GB labels has a more significant decrease. We believe that under the blessing of causality between stages, unlabeled data could better remedy the sample bias and improve the model generalization ability.

Finally, the advantage of the auxiliary tasks is explored. The MSIS consists of 3 business stages that are the AR, WS and GB stage with the first 2 as auxiliary tasks while the last as primary task. We define one task case as keeping only the AR stage for auxiliary task. As shown in the Fig. \ref{ablation}(b), the AUC of the two task case on the 3 GB labels is higher than that of the one task case. It is important to maintain complete causal relationship corresponding to business processes.

\begin{figure*}[htbp]
\centering
\subfigure[ Embedding dimension ]{\label{}\includegraphics[scale=0.45]{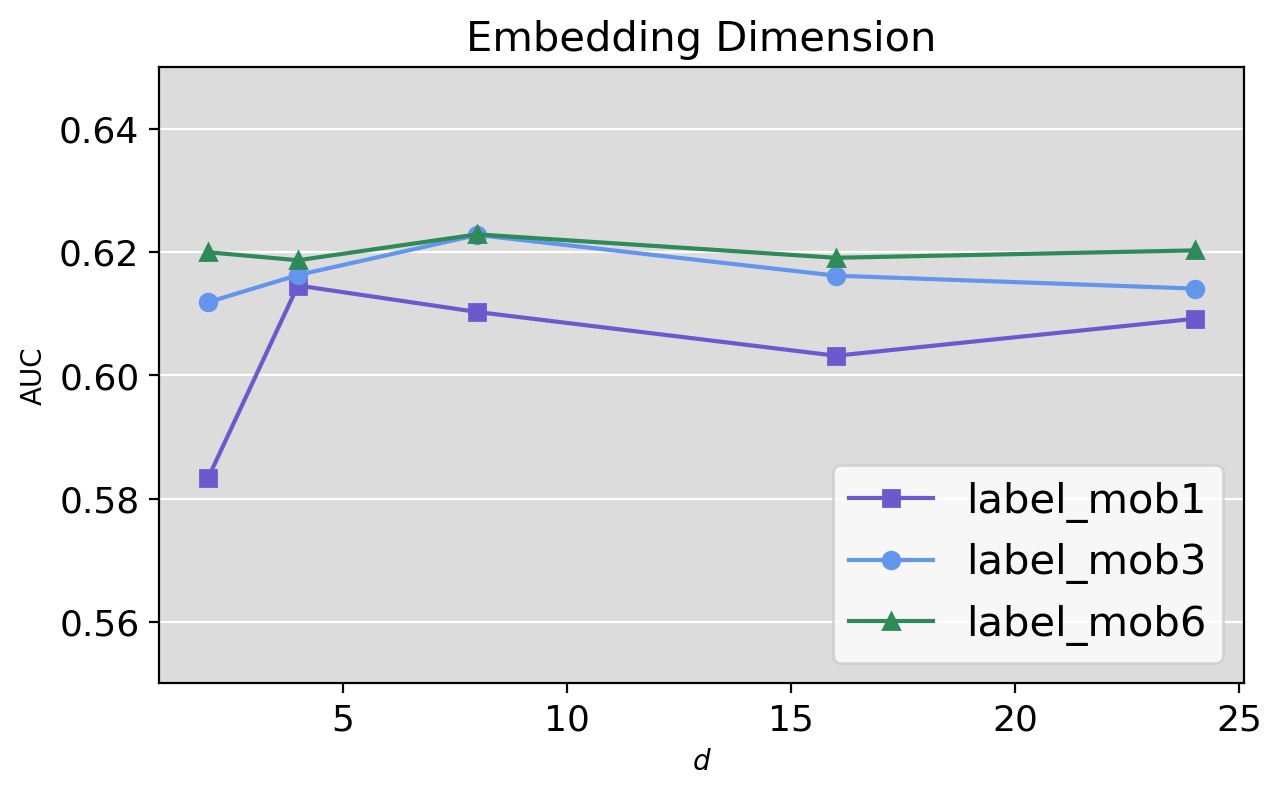}}
\subfigure[ Semi-supervised loss weight ]{\label{}\includegraphics[scale=0.45]{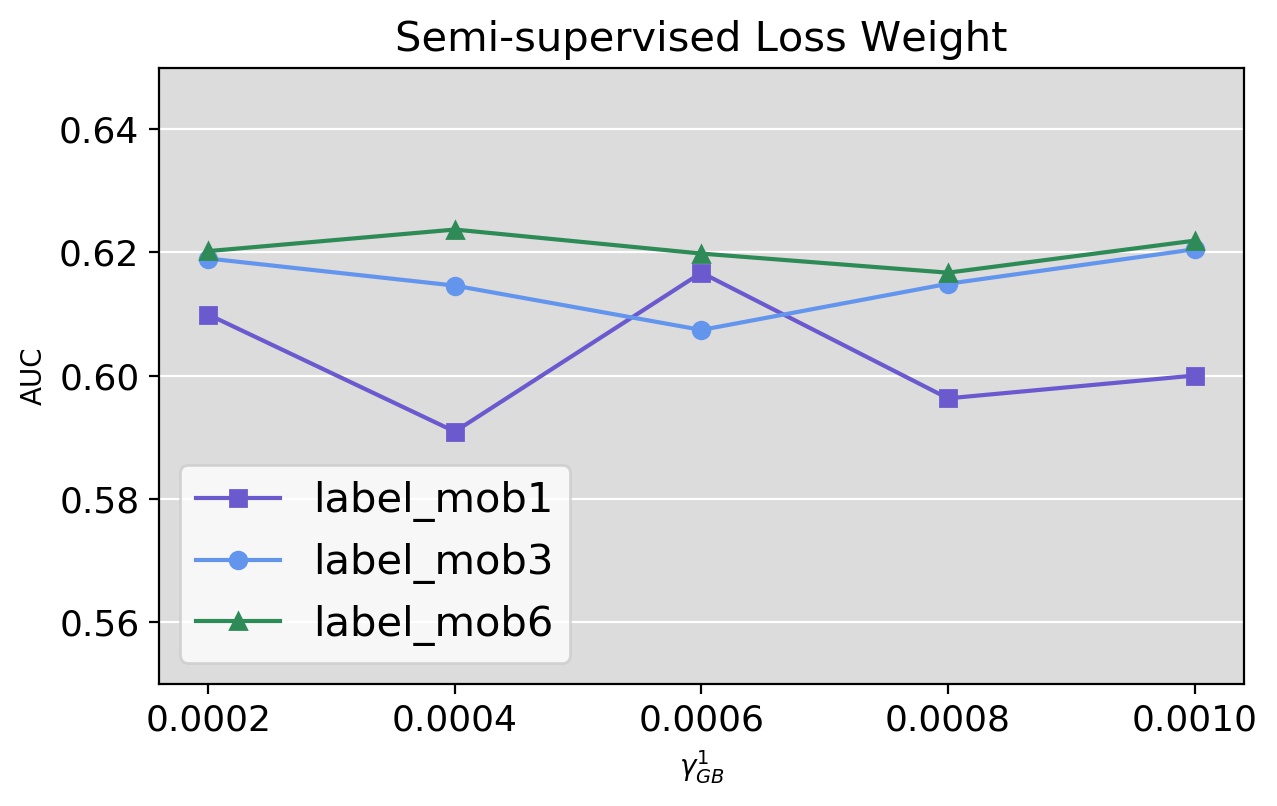}}
\caption{The mean AUC performance in Parameter Study }
\label{hyper}
\end{figure*}

\subsection{Parameter Study}
The parameter sensitivity is also explored.

\textbf{embedding dimension}: we vary the embedding dimension as $[2,4,8,16,24]$, the results are shown in Fig. \ref{hyper}(a). When the embedding dimension is too low and only has 2 dimensions, the AUC of the model in $label\_mob1$ decreases. The model is stable on other dimensions, indicating that MSIS is not very sensitive to the embedding dimension parameter. When the embedding dimension is small, the model only learns less knowledge, and the fitting degree of the data distribution is not enough. Usually, with the increase of the embedding dimension, the model effect will first increase, then saturate, and then decrease. Our model also roughly conforms to this trend. Synthetically, the AUC on the three labels is the highest when $d=8$. So, we finally set $d=8$ as the embedding dimension in all experiments.

\textbf{semi-supervised loss weight}: Because the total number of labeled data and unlabeled data is quite different, the training balance between them is quite important for the network performance. Since unlabeled samples account for the majority, the value of weight is relatively small. As shown in the Fig. \ref{hyper}(b), the results of $label\_mob3$ and $label\_mob6$ under different weights are stable. However, $label\_mob1$ fluctuates to a certain extent, the reason may be that the imbalance of positive and negative samples of this label is more serious. Finally, we choose $\gamma_{GB}=6e-4$ as the weight parameter in experiments.

\section{Conclusion And Future Work}
In this paper, we propose a MSIS method capturing the sequential pattern of interactions among multiple stages of loan business to make better use of the underlying causal relationship when modeling credit. Specifically, we first define 3 stages with sequential dependence throughout the loan process and integrate them into a multi-task architecture. Inside stage, a intra-stage multi-task classification is built for different business goals. Then we design an Information Corridor to express sequential dependence, leveraging the information from former stages via a hierarchical attention module. In addition, semi-supervised loss is introduced to deal with the unobserved customers. Experimental results on a real data set show the ability to remedy the population bias and improve model generalization ability.

Classification is a common method in the practice of credit model, as all tasks defined in this work are binary classification. While, in addition to identifying user as good or bad, we can also define some relative relationships from the perspective of business experience. Relationships defined by business experience can be either explicit or ambiguous. To better integrate models with business experience, the idea of weakly supervised learning and ranking loss are the directions of our future exploration.

\bibliographystyle{IEEEtran}
\bibliography{bibfile}

\end{document}